# Topological surface state dominated nonlinear transverse response and microwave rectification at room temperature

Qia Shen,[1] Jiaxin Chen,[1] Bin Rong,[2] Yaqi Rong,[2] Hongliang Chen,[1] Tieyang Zhao,[4] Xianfa Duan,[3] Dandan Guan,[1,6,7] Shiyong Wang,[1,6,7] Yaoyi Li,[1,6,7] Hao Zheng,[1,6,7] Xiaoxue Liu,[1,6,7] Xuepeng Qiu,[5] Jingsheng Chen,[4] Longqing Cong,[5] Tingxin Li,[1,6,7] Ruidan Zhong,[1] Canhua Liu,[1,6,7] Yumeng Yang,[2†] Liang Liu,[1,6,7‡] Jinfeng Jia[1,6,7†]

[1]*Tsung-Dao Lee Institute, Shanghai Jiao Tong University, Shanghai, 201210, China*
[2]*Shanghai Engineering Research Center of Energy Efficient and Custom AI IC, School of Information Science and Technology, ShanghaiTech University, Shanghai 201210, China*
[3]*Shanghai Key Laboratory of Special Artificial Microstructure Materials and Technology and School of Physics Science and Engineering, Tongji University, Shanghai, 200092, China*
[4]*Department of Materials Science and Engineering, National University of Singapore, 117575, Singapore*
[5] *State Key Laboratory of Optical Fiber and Cable Manufacture Technology, Department of Electrical and Electronic Engineering, Southern University of Science and Technology, Shenzhen, 518055 China*
[6] *Key Laboratory of Artificial Structures and Quantum Control (Ministry of Education), School of Physics and Astronomy, Shanghai Jiao Tong University, Shanghai 200240, China*
[7] *Hefei National Laboratory, Hefei 230088, China*

[†]Contact author: yangym1@shanghaitech.edu.cn
[‡]Contact author: liul21@sjtu.edu.cn
[†]Contact author: jfjia@sjtu.edu.cn

**ABSTRACT**. Nonlinear Hall effect (NLHE) offers a novel means of uncovering symmetry and topological properties in quantum materials, holding promise for exotic (opto)electronic applications such as microwave rectification and THz detection. Generally, NLHE can be induced by Berry curvature dipole (BCD), skew scattering, and side jump. In trigonal or hexagonal crystals, the three-fold rotational symmetry forbids BCD but may allow for skew scattering if inversion symmetry is broken in the bulk or the surface. This BCD-independent NLHE could exhibit a robust response even at room temperature, which is highly desirable for practical applications. However, in materials with bulk inversion symmetry, the coexistence of bulk and surface conducting channels often leads to a suppressed NLHE and complex thickness-dependent behavior. Here, we report the observation of room-temperature nonlinear transverse response in 3D topological insulator $Bi_2Te_3$ thin films, whose electrical transport properties are dominated by topological surface state (TSS). By varying the thickness of $Bi_2Te_3$ epitaxial films from 7 nm to 50 nm, we found that the nonlinear transverse response increases with thickness from 7 nm to 25 nm and remains almost constant above 25 nm. This is consistent with the thickness-dependent basic transport properties, including conductance, carrier density, and mobility, indicating a pure and robust TSS-dominated linear and nonlinear transport in thick (>25 nm) $Bi_2Te_3$ films. The weaker nonlinear transverse response in $Bi_2Te_3$ below 25 nm was attributed to Te deficiency and poorer crystallinity. By utilizing the TSS-dominated electrical second harmonic generation, we successfully achieved the microwave rectification from 0.01 to 16.6 GHz in 30 nm and bulk $Bi_2Te_3$. Our work demonstrated the room temperature nonlinear transverse response in a paradigm topological insulator, addressing the tunability of the topological second harmonic response by thickness engineering.

## I. INTRODUCTION.

Hall effect and its family occupy a prominent position in condensed matter physics [1-4]. In systems with time-reversal symmetry, the Onsager relation [5] typically prohibits the occurrence of a linear charge Hall response. In 2015, Sodemann and Fu proposed a new type of Hall effect, namely the nonlinear Hall effect (NLHE), which can be allowed in non-magnetic materials with inversion-symmetry breaking [6]. The NLHE is proportional to the Berry curvature dipole (BCD), which could exist in 2D transition metal dichalcogenides [7-10], 3D Weyl semimetal [11], Moiré superlattice [12], etc. In these systems, there is usually no more than one mirror plane, which means only materials with limited point groups (eg., $C_1$, $C_{1v}$) can have this effect [13]. In addition to the BCD, skew scattering and side-jump scattering, were also suggested to produce nonlinear transverse response [12, 14], which can exist in materials with higher symmetries (such as $C_3$, $C_{3v}$, $C_{3h}$, $D_3$, $D_{3h}$), as observed in some topological materials [15, 16] and elemental materials [17, 18]. In these trigonal and hexagonal systems, the BCD is not allowed by symmetry, while the skew scattering produced by the inherent spin chirality on the Fermi surface[15] in these systems may give rise to a sizable NLHE due to the inversion symmetry breaking at the surface state or the bulk state. Though this BCD-independent NLHE has been found to exist strongly from 2D to 3D limit, the entanglement of the surface and bulk conducting channel generally leads to a suppressed response mainly due to the shunting effect. A clear surface state-dominated nonlinear transport behavior in a single material is still lacking. More importantly, the nonlinear transport behaviors [15, 19] reported in 3D topological insulators are restricted to low temperatures (≤ 200 K), which prevents their practical applications. Among all topological insulators, $Bi_2Te_3$ possesses robust TSS with featured hexagonal warping even at room temperature [20-22]. Therefore, the room-temperature electrical second harmonic response of $Bi_2Te_3$ and its applications in optoelectronic devices using its topological band structure, are well worth investigating.

In this work, we report the TSS-dominated room-temperature nonlinear transverse transport in $Bi_2Te_3$ thin films. By preparing $Bi_2Te_3$ with different thicknesses from 7 nm to 50 nm, we found that both the linear and nonlinear transport behaviors are thickness independent when $Bi_2Te_3$ is thicker than 25 nm, revealing a TSS-dominated regime with high sample quality. Furthermore, we demonstrated the GHz wave rectification in a 30 nm $Bi_2Te_3$ film and a bulk $Bi_2Te_3$ crystal, highlighting their potential in future nonlinear electronics applications.

## II. RESULTS

### A. Epitaxial growth, structural characterization and basic transport of $Bi_2Te_3$ thin films

We grew $Bi_2Te_3$ films on atomically flat $Al_2O_3$ (0001) substrates through molecular beam epitaxy (MBE) technique (see Methods). As illustrated in Fig. 1(a), one quintuple layer (QL) of $Bi_2Te_3$ consists of 5 atomic layers, *i.e.,* Te-Bi-Te-Bi-Te, as confirmed by the high-angle annular dark field-scanning transmission electron microscopy (HADDF-STEM) image [Fig. 1(d)]. The 3 sharp peaks shown in the high-resolution X-ray diffraction phi-scan pattern [Fig. 1(b)] reveals that the $Bi_2Te_3$ (0001) film possesses a twin-free crystalline structure, consistent with the three-fold rotational symmetry ($C_{3v}$) of typical hexagonal structure, which was also a prerequisite to probe the symmetry-dependent nonlinear transport in 3D topological insulators [15, 19]. The good epitaxy and atomically flat surface morphologies of $Bi_2Te_3$ thin films are further verified through atom force microscope (AFM) imaging [Supplemental Material, Fig. S1(c)] and streaky pattern [Supplemental Material, Fig. S1(d)] observed by in-situ reflection high-energy electron diffraction (RHEED). More importantly, as shown in Supplementary Material Fig. S1(c), most of these triangles are aligned along the same crystal orientation, further confirming the twin-free property of the film. The high quality of $Bi_2Te_3$ is a necessity of realizing a dominating topological surface state over the bulk state [21, 23, 24] and is also important for studying the thickness-independent transport behavior [25, 26].

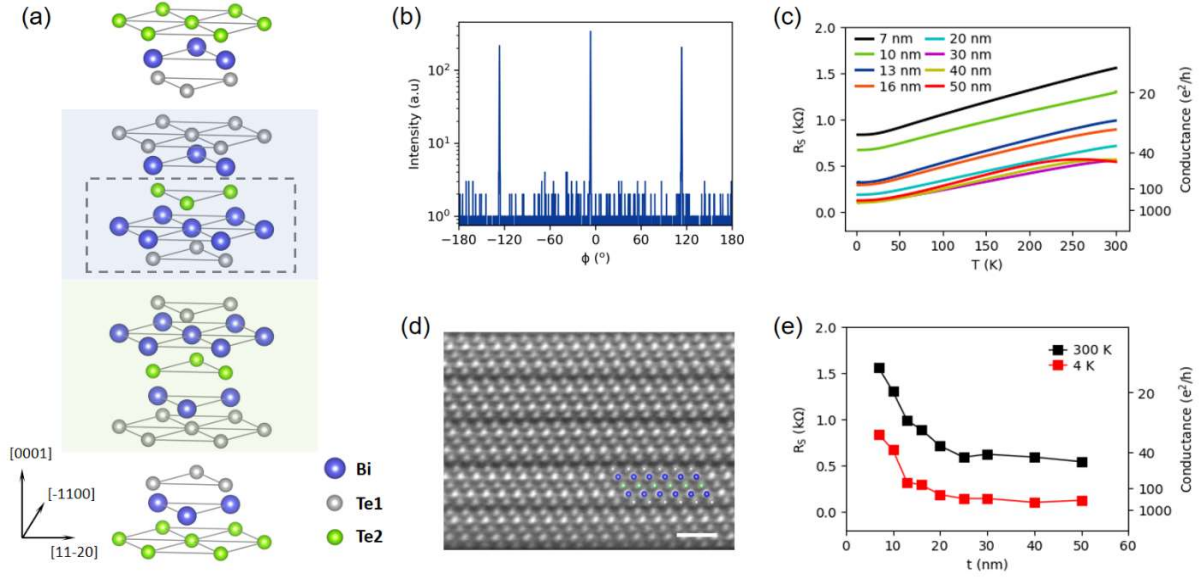

FIG. 1. Crystal structure and resistivity of $Bi_2Te_3$ thin films. (a) Crystal structure of $Bi_2Te_3$. The green shadow outlines a quintuple layer which is an inversion patterner to its adjacent quintuple layer (shadowed by blue). The three atomic layers Te/Bi/Te outlined by the dashed rectangle were replotted in Fig. 2(a). (b) High-resolution X-ray diffraction phi-scan pattern with $Bi_2Te_3$ (113) plane rotated along [111] axis. (c) Temperature dependence of sheet resistance of $Bi_2Te_3$ films with thickness ranging from 7 to 50 nm. (d) The high-angle annular dark field (HAADF) scanning transmission electron microscopy (STEM) image of $Bi_2Te_3$ grown on $Al_2O_3$ (0001). The scale bar is 1 nm. (e) Thickness dependence of sheet resistance for $Bi_2Te_3$ films at 4 K (red) and 300 K (dark).

Figure 1(c) shows the temperature ($T$) dependence of the sheet resistance ($R_s$) of 7 nm to 50 nm $Bi_2Te_3$, all exhibiting metallic behavior. The thickness ($t$) dependences of $R_s$ at 300 K and 4 K are plotted in Fig. 1(e). We observed that $R_s$ firstly decreases with increasing thickness from 7 nm to 25 nm and then keeps almost constant for thicknesses above 25 nm. This behavior is similar to that reported in $Bi_2Te_3$ films grown on $BaF_2$ substrate [24]. The nearly unchanged $R_s$ with $t$ suggests a dominating TSS (revealed by the conductance values) over the bulk state in 25-50 nm $Bi_2Te_3$ [24, 25]. While for $Bi_2Te_3$ blow 25 nm, imperfect crystallinity may degrade the TSS, resulting in increased resistivities [24]. We also measured the ordinary Hall effect at 300 K for these samples as shown in Fig. 2(b). The nearly linear $R_{xy}$-$H_z$ characteristic indicates a single type of carrier from the surface state for all thicknesses and the negative Hall coefficient ($R_H = \frac{R_{yx}}{H}$) indicates that the transport is in the N-type regime and Te deficiency effect [26, 27]. In addition, the nearly constant Hall coefficient for $Bi_2Te_3$ above 16 nm is almost the same (Supplemental Material, Fig. S3), which indicates the dominance of TSS with an insulating bulk state, consistent with the result of resistance measurement in Fig. 1(e).

### B. Thickness dependence of nonlinear transverse response at room temperature

For the nonlinear transport measurements, we fabricated Hall bar devices with different angles ($\phi_I$) between current path and [11-20] direction, as shown in Fig. 2(a) (see Methods for the details). The plane view of the (0001) hexagon has a three-fold rotational symmetry with corresponding mirrors. We denote [11-20] (and equivalent crystalline orientation) as the "low-symmetry" axis since the mirror symmetry is broken with respect to the plane defined by [11-20] and [0001]. Similarly, [1-100] is denoted as the "high symmetry" axis. Figure 2(c) and 2(d) show the nonlinear Hall resistance $R_{2\omega}$ as a function of the ac current amplitude ($I_{ac}$) for four typical $\phi_I$ angles. We found that the nonlinear Hall signals exist when the current flows along the low symmetry directions ($\phi_I = 0°$ and $180°$) and $R_{2\omega}$ scales linearly with $I_{ac}$, representing a typical feature for nonlinear transverse response, but the nonlinear signals vanish along the high symmetry directions ($\phi_I = 30°$ and $210°$) in Fig. 2(d). According to the linear fit, we summarized the slope ($R_{2\omega}/I_{ac}$) as a function of $\phi_I$ for different $Bi_2Te_3$ thicknesses in Fig. 2(e). The $R_{2\omega}/I_{ac}$-$\phi_I$ relation for all thicknesses clearly reveals threefold rotational symmetry, which corresponds to the crystal symmetry of the $Bi_2Te_3$ structure.

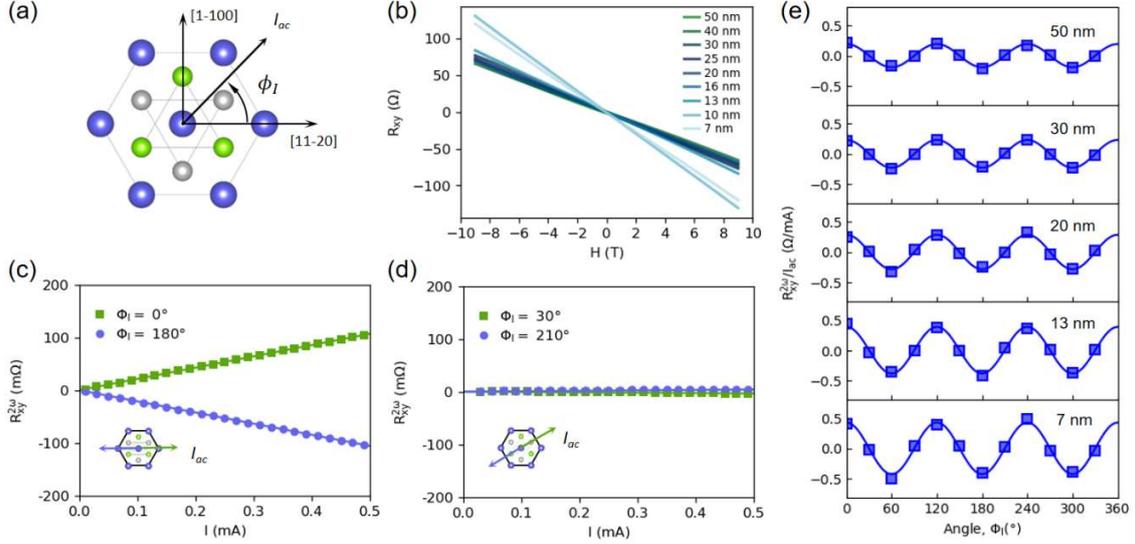

FIG. 2. Room temperature ordinary Hall effect and nonlinear transport in $Bi_2Te_3$ thin films. (a) The definition of the azimuth angle ($\phi_I$) of the current with respect to the [11-20] direction of the (0001) hexagonal. (b) Magnetic field dependence of the Hall resistance of $Bi_2Te_3$ films with varying thicknesses at 300 K. (c) (d) Second harmonic Hall response as a function of ac current for 30 nm $Bi_2Te_3$ along low symmetry directions $\phi_I = 0°$ and $180°$ (c), and high symmetry directions $\phi_I = 30°$ and $210°$ (d). The current flowing direction is indicated by green and blue arrows in the insets. (e) Current angle dependence of $R_{2\omega}/I$ for different $Bi_2Te_3$ thicknesses.

### C. TSS-dominated electrical second harmonic generation and its mechanism

To characterize the strength of the nonlinearity, one can define the NLHE susceptibility $\frac{E_y^{2\omega}}{(E_x^\omega)^2}$, where $E_y^{2\omega} = \frac{V_y^{2\omega}}{W}$, and $E_x^\omega = \frac{V_x^\omega}{L} = \frac{I_{ac}R_{xx}}{L}$ are the second harmonic transverse and first harmonic longitudinal electric fields respectively ($L$ and $W$ are the length and width of the Hall bar device, respectively). $V_y^{2\omega}$ is the second order transverse voltage and $R_{xx}$ is the normal resistance. By measuring $V_y^{2\omega}$ of $Bi_2Te_3$ devices (see Methods for details), we can obtain $\frac{R_{xy}^{2\omega}}{I_{ac}}$, from which we calculate $\frac{E_y^{2\omega}}{(E_x^\omega)^2}$ and its thickness dependence is shown in Fig. 3(a). We found that $\frac{E_y^{2\omega}}{(E_x^\omega)^2}$ initially increases for $t < 25$ nm, and then remains nearly unchanged when $t$ is above 25 nm, which corresponds to a TSS-dominated region [the gray region in Fig.3(a)].

In previous studies, the thickness dependence of NLHE has rarely been studied, partially due to the mixture of the surface and bulk state and the difficulty of device fabrication, especially for 2D materials with systematical thicknesses. For example, the BCD-induced NLHE, as in $WTe_2$ [8] or $TaIrTe_4$ [28], can only originate from the surface state to meet the requirement of inversion symmetry breaking while the bulk state is non-insulating. In this case, the NLHE should decrease with increasing thickness. For a recent investigation on BiTeBr [29] with $C_{3v}$ symmetry, the inversion symmetry is broken in both the surface and the bulk, concurrently giving rise to a Rashba spin-orbit interactions induced skew scattering and NLHE. As a result, a non-monoclinic thickness dependence behavior was observed. In fact, for both $TaIrTe_4$ and BiTeBr, the NLHE responses were only compared among three thicknesses, which only provide limited experimental indication on the thickness dependent behavior and the dominated nature (bulk or surface) of the effect. Here for $Bi_2Te_3$, the nonlinear susceptibility increases with thickness and then saturate to $6.2\times10^{-3}$ μm/V [the dash line in Fig. 3(a)], which differs markedly from the reported behaviors in other systems. To understand this phenomenon, we check in details the crystal quality for the $Bi_2Te_3$ films. We measured θ-2θ XRD scan [Supplemental Material, Fig. S2(a)], and found the diffraction peaks of (0006), (0015), (0018) and (0021) become sharper with increasing thickness, indicating a better crystal quality for thicker films. According to the Lorentzian fit of the (0015) peaks, the coherent scattering length ($D_z$), characterizing the average grain size, is drawn in Fig. 3(b). $D_z$ increases with thickness and implies the higher quality (crystallinity) for thicker films. Meanwhile, the Te deficiency effect for thinner $Bi_2Te_3$ films is more prominent than that in thicker ones, as discussed above through the ordinary Hall measurements and $R$-$T$ curves. Therefore, as schematically plotted in the left insert of Fig.3(a), the imperfect crystallinity could make degradation effects [30-33] on TSS, and the bulk conductive band (BCB) would be below the $E_F$ with the Dirac point buried in the bulk valence band (BVB). While for $Bi_2Te_3$ above 25 nm [the right insert of Fig.3(a)], the BCB above $E_F$ because of band bending [26, 34] and the higher quality of the films jointly lead to dominated TSS. We also checked carrier mobility ($\mu$) at room temperature according to the normal resistance and Hall resistance measurement in Figs. 1(e) and 2(b), respectively. We found that the mobility also increases with the thickness and saturates to around 130 $cm^2V^{-1}s^{-1}$, which is marginally higher than the previous work [35, 36], further revealing the good sample quality in thick $Bi_2Te_3$. By comparing Figs. 3(a-c), a very similar tendency on thickness for $\frac{E_y^{2\omega}}{(E_x^\omega)^2}$, $D_z$ and $\mu$ is observed. Therefore, the enhancement of the electrical second harmonic generation in thick $Bi_2Te_3$ is strongly correlated to the crystallinity and/or quality of the material. When $t$ is small, the imperfect crystallinity results in a degraded TSS and thus a weak response. As $t$ increases, the growth and annealing time increase, contributing to the refined crystallinity of $Bi_2Te_3$, which could be revealed by XRD and mobility results. As a result, a robust TSS could exist, giving rise to an enhanced NLHE. In fact, a nearly pure TSS state with an insulating bulk state has already been achieved in a large thickness range, eg. from 8 QL to 256 QL for $Bi_2Se_3$ [23] and from 10 QL to 50 QL for $Bi_2Te_3$ [24]. Ideally, the thickness dependence of the NLHE in our $Bi_2Te_3$ can directly correlate with the strength of its TSS, which is the only source to provide the inherent chirality of the Bloch electrons. Therefore, the second harmonic signal directly reflects the strength of a nearly pure TSS instead of a mixture of surface and bulk state.

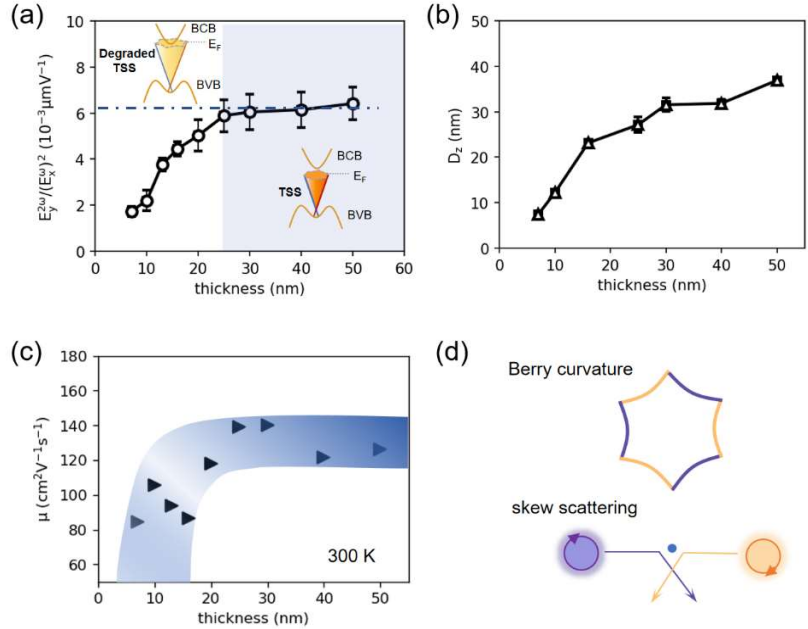

FIG. 3. Room temperature electrical second harmonic generation in $Bi_2Te_3$ thin films. (a) Thickness dependence of $E_y^{2\omega}/(E_x^\omega)^2$. The error bars come from the standard derivation of the measured values for three low-symmetry axes. The gray region is the TSS-dominated region with the dashed line presenting the maximum of $E_y^{2\omega}/(E_x^\omega)^2$. The two schematic band structures inserted represent a band bending from the film <25 nm (left) to ≥25 nm (right). (b) The estimated coherent scattering length ($D_z$) as a function of the thickness of $Bi_2Te_3$ film. (c) The mobility of $Bi_2Te_3$ films with different thicknesses. (d) Berry curvature distribution of topological insulator $Bi_2Te_3$, and the schematic diagram of the skew scattering of chiral surface Dirac fermion inducing nonlinear transport.

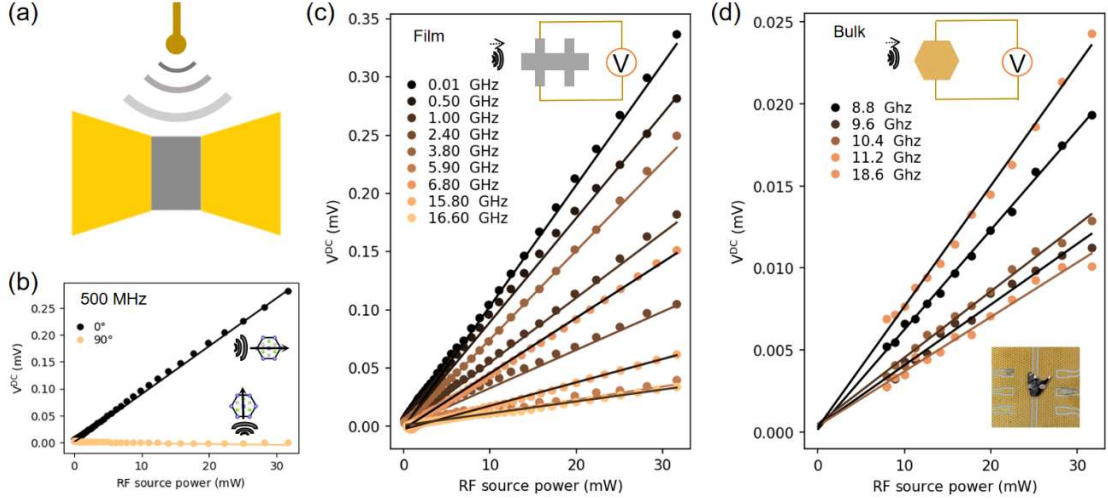

FIG. 4. Room-temperature RF rectification of $Bi_2Te_3$. (a) The schematic figure of a rectifier. The rectified DC response can be detected transverse to the incident electric field. (b) Hall voltage measured as RF source power changes, with $I_{RF}$ flows in 0° (low symmetric direction) or 90° (high symmetric direction) at 500 MHz. (c), (d) The rectified DC electrical transverse voltage as a function of the RF power measured at various frequencies, with microwave current in the low symmetric direction, for 30 nm film (c) and bulk (d). The schematic and physical figures are inserted.

With the above arguments, we discuss the microscopic origin of the NLHE for our $Bi_2Te_3$ in more detail. Theoretically, $Bi_2Te_3$ possesses TSS with featured hexagonal warping [37] as shown in Fig. 4(a), and its robustness at room temperature has already been demonstrated by angle-resolved photoemission spectroscopy (ARPES) experiments [20-22]. Typically, the band structure of $Bi_2Te_3$ is linearly dispersed near the Dirac cone and the trigonal crystal symmetry will give rise to a prominent hexagonal warping effect due to the cubic wave vector term [37] [see the inserts in Fig. 3(a) and see Supplemental Material for the details]. And with thickness increasing, the Fermi level $E_F$ shifts downward (eg., ~ 40 meV from 10 nm to 50 nm [24] and ~70 meV from 16 nm to 70 nm [26] in previous work), which promotes the purity of the TSS and may also affect the NLHE [see Insets in Fig. 3(a)]. It is noted that typically the energy distance from $E_F$ to Dirac point for $Bi_2Te_3$ film is approximately 100~250 meV, while the bottom of bulk conduction band is usually higher [20-22, 27]. The schematic diagram of Berry curvature distribution of hexagonally warped Fermi surface for $Bi_2Te_3$ is shown in Fig. 3(d), which gets distributed into 6 segments, positive and negative, from which a chiral property is revealed. For $Bi_2Te_3$, because of $C_{3v}$ symmetry, similar with other threefold symmetry materials, such as $Bi_2Se_3$ [15], $CoTe_2$ [38] and Te [18], the BCD could not be allowed. In this case, only the skew scattering and side jump scattering need to be considered, while the former dominates because of weak impurity limit ($\tau \rightarrow \infty$) for its high-order $\tau$ dependence, where $\tau$ is the electron scattering time [15, 39]. As schematically drawn in Fig. 3(d), the finite second harmonic generation originates from the self-rotation of the wave packet from states on the Fermi surface due to finite Berry curvature, and even an isotropic scatterer deflects the motion of wave packets in a preferred direction [15, 39]. Therefore, the physical origin of the NLHE in $Bi_2Te_3$ could also be the skew scattering. However, a typical scaling between the NLHE susceptibility and the quadratically bulk conductivity ($\sigma_b^2$) as shown in previous reports could not be simply made here for $Bi_2Te_3$, since there is no existing theory on the proper scaling between the NLHE susceptibility and the surface conductivity ($\sigma_s$) in a surface-state-dominated regime.

### D. Radiofrequency (RF) rectification experiment

TABLE I. The comparison of different materials.

| Materials | $T_{max}$ (K) | $E_y^{2\omega}/(E_x^\omega)^2$ $10^{-3}$ μm V$^{-1}$ | Preparation method | RF test range (GHz) | Symmetry | Origin |
|---|---|---|---|---|---|---|
| Bilayer WTe$_2$ [7] | 100 | / | Mechanical exfoliation | None | $C_s$ | BCD |
| Few-layer T$_d$-WTe$_2$ [8] | 100 | ≈ 0.6 | Mechanical exfoliation | None | $Pm$ | BCD, scattering |
| Twist WSe$_2$ [9] | 30 | ≈ 4×10$^3$ | Mechanical exfoliation | None | $C_1$ | BCD |
| TaIrTe$_4$ [28] | 300 | -5 (16 nm) | Mechanical exfoliation | ~2.4 | $1m$ | BCD |
| BaMnSb$_2$ [40] | 400 | ≈ 2×10$^5$ (300 K) | Micro flakes | 0.3 | $C_{2v}$ | BCD |
| BiTeBr [29] | > 350 | ≈ 500 (300 K) | Mechanical exfoliation | 1~5.9 | $C_{3v}$ | Skew scattering |
| Bi$_2$Se$_3$ [15] | 200 | ≈ 0.42 (20 nm) | Epitaxial growth | None | $C_{3v}$ | Skew scattering |
| Bi$_2$Te$_3$ (this work) | > 300 | 6 (30 nm, 300 K) | Epitaxial growth | 0.01~16.6 | $C_{3v}$ | Skew scattering |

Furthermore, as a demonstration for the NLHE application, we performed the microwave rectification ("frequency doubling" in some contexts) in thick-film and bulk Bi$_2$Te$_3$ in the following. Figure 4(a) shows the schematic of the microwave rectification for Bi$_2$Te$_3$. Under the exposure to the microwave, the RF current $I_{RF}$ passes through Bi$_2$Te$_3$, resulting in a rectified DC current $J_{DC}$ transverse to the incident electric field $E_\omega$ due to the NLHE. As shown in Fig. 4(c), for a wide RF frequency range from 0.01 to 16.6 GHz, the rectified voltage $V^{DC}$ of 30 nm Bi$_2$Te$_3$ presents a good linear relationship with the power of the RF source ($\propto I_{RF}^2$) as shown in Fig. 4(c), indicating a typical second-order nonlinear response. Importantly, the $V^{DC}$ can be observed for low symmetry axes ($\phi_I = 0°$) while it vanishes along high symmetry axes ($\phi_I = 90°$) at 500 MHz [Fig. 4(b)], which is consistent with the low-frequency (17.7-777.77 Hz) results and confirms a crystal symmetry protected behavior for this effect. Interestingly, we also explore the RF rectification on bulk Bi$_2$Te$_3$ and a sizable $V^{DC}$ can still be observed, though the imperfect contact between our bulk Bi$_2$Te$_3$ crystal and the waveguide plane [see the optical image in Fig. 4(d)]. These results indicate that the second order nonlinear response in Bi$_2$Te$_3$ persists in a large frequency range from 30 nm to bulk. There are many other materials where NLHE exists, including 2D TMD materials [7-9], Weyl semimetals [28], non-centrosymmetric Dirac materials [40], 2D Rashba materials [29]. For comparison, we provided the NLHE related parameters, including susceptibility, growth method, crystal symmetry, tested temperature, etc., in Table I. First of all, most materials like 2D TMD [8-10] and graphene [12] have low working temperatures, and the signals will vanish as temperature increases, leading to great limitations for room temperature applications. In contrast, TaIrTe$_4$ [28], BiTeBr [29], BaMnSb$_2$ [40] and Bi$_2$Te$_3$ are able to reach high working temperatures (>300 K). For 30 nm Bi$_2$Te$_3$ in our study, an NLHE susceptibility of 6×10$^{-3}$ μm V$^{-1}$ is observed at 300 K, which is comparable to that of 16 nm TaIrTe$_4$ [28]. Robust and dominated TSS is the main reason for the skew-scattering induced NLHE survived in Bi$_2$Te$_3$ at room temperature. Furthermore, as epitaxial films, Bi$_2$Te$_3$ and Bi$_2$Se$_3$ [15] exhibit superior scalability and processability than those ultra-thin flakes obtained by mechanical exfoliation or micro flakes using flux method, which has been studied a lot in the field of NLHE (See Table I). More importantly, as a 3D topological insulator, Bi$_2$Te$_3$ will not be restricted by thickness because of its conductive TSS and insulating bulk, which means it may overcome the thickness limit of NLHE, similar to BaMnSb$_2$. As compared in Table I, among typical NLHE materials, few of them can perform at a high frequency above 10 GHz. The good frequency response behavior reveals the potential of Bi$_2$Te$_3$ as a broadband microwave rectifier and energy harvester.

### III. CONCLUSIONS

We discovered the nonlinear transverse response in different thick Bi$_2$Te$_3$ under zero magnetic field, finding that in the thicker films the TSS dominated and further improved the nonlinear response at room temperature. Then we completed RF rectification on the Bi$_2$Te$_3$ film and bulk sample with varying frequencies up to 16.6 GHz. The NLHE mechanism BCD is sensitive to the band structure, and always requires low symmetry to break inversion symmetry, which is strict for materials and unsuitable for most surfaces or interfaces. Whereas skew scattering can apply to a wider class of materials with inversion symmetry, triangular and hexagonal systems for instance (see Table I), which account for a large proportion in nature. Moreover, skew scattering induced nonlinear transverse response could be tuned by intrinsic TSS, promising to expand the nonlinear transport frequency regime from low frequency to GHz and THz, so it could realize microwave rectification and further realize THz detection [39]. Unlike some other systems that benefit from surface or interface effect, or integrate conductive bulk states, $Bi_2Te_3$, as a 3D topological insulator with an insulating bulk would advance the development of room-temperature electrical second harmonic generation without limitations on thickness.

## IV. METHODS

### A. Sample growth and device fabrication

After ultrasonic immersion for 5 min in acetone and ethanol separately, $Al_2O_3$ (0001) substrates (HF-Kejing, China) are annealed in air at 1100 °C for 1 h. Then the substrates are degassed in ultra-high vacuum MBE chamber at 650 °C for 1 h. Finally, high-quality $Bi_2Te_3$ films are grown on $Al_2O_3$ substrates by MBE technique with base system pressure ＜$1\times10^{-10}$ Torr and growth pressure about $1\times10^{-9}$ Torr. Under a Te/Bi flux ratio of ~20, Te (99.9999%) is evaporated for 30 s to make Te-rich environment before Bi (99.999%) and Te are evaporated together from standard Knudsen cells while the substrates stay at 230 °C. After growth, the films are annealed at the same temperature with rich Te for 5 min and then without Te for 5 min to get higher quality. The $Bi_2Te_3$ films we grow are of good epitaxy with a clear step of ~1 nm which is equal to 1 QL. Our samples of different thicknesses from 7 to 50 nm are fabricated to hall bar devices by laser direct writing photolithography and argon ion milling.

### B. STEM Sample Preparation and Characterization

The STEM sample of the $Bi_2Te_3$/ $Al_2O_3$ (0001) was fabricated by a focused ion beam (FIB) machine (ZEISS Crossbeam 350) using a Ga ion beam. STEM characterization was conducted by Spectra 300 STEM (ThermoFisher Scientific) equipped with the powerful combination of X-CFEG and the S-CORR probe aberration corrector operating at an accelerating voltage of 300 kV. The HAADF-STEM images were acquired with a probe forming angle of 30 mrad and a collection angle of 46-200 mrad.

### C. Nonlinear transverse transport measurement

The basic transport properties of the devices were measured in Quantum Design Physical Property Measurement System (PPMS). Second-harmonic voltage signal was gathered by lock-in amplifiers (Stanford Research System SR830) with a harmonic current applied to the device by Keithley 6221 current sources. The phase of the second-harmonic voltages was ~ 90° during the measurement.

### D. RF rectification measurement

Vector network analyzer N5227A (Keysight) was used to generate RF signals of frequencies ranging from 0.01 to 20 GHz and apply RF signal to the $Bi_2Te_3$ device. A Keithley 2182 nanovoltmeter was used to measure the output rectified DC voltage.

## ACKNOWLEDGMENTS

We thank the helpful discussions with Pan He, Peng Chen and Liang Fu. The research was supported by the Ministry of Science and Technology of China (Grant Nos. 2019YFA0308600, 2021YFA1401500 and 2020YFA0309000), the National Natural Science Foundation of China (Grant Nos. 12474121, 11861161003, 12104293, 92365302, 22325203, 92265105, 92065201, 12074247 and 12174252), the Strategic Priority Research Program of the Chinese Academy of Sciences (Grant No. XDB28000000), the Science and Technology Commission of Shanghai Municipality (Grant Nos. 19JC1412701, 2019SHZDZX01 and 20QA1405100), the Innovation Program for Quantum Science and Technology (Grant No. 2021ZD0302500) and the China National Postdoctoral Program for Innovative Talents (Grant No. BX2021185). L.L. acknowledges the Yangyang Development Fund and the Xiaomi Young Scholar program. This work was supported by the National Center for High-Level Talent Training in Mathematics, Physics, Chemistry and Biology.

L.L. and Q.S. conceived the experiments. Q.S., J.C., and B.R. performed thin film deposition and device fabrication. Q.S. carried out transport measurements and data analysis. Y.R. and Q.S. completed the RF reflection measurement. H.C. contributed to device fabrication. X.D. contributed to thin film deposition. Q.S. did the AFM characterization, XRD, and STEM samples preparation using FIB. T.Z. did the Φ scan. R.Z. grew the $Bi_2Te_3$ bulk crystal. The other authors contributed to data analysis. L.L., Y.Y., J.J. and Q.S. wrote the manuscript and all authors contributed to its final version.

The authors declare no competing financial interests.

# Supplemental Material

## 1. The characterization of Bi$_2$Te$_3$ and the substrate

As shown in Fig. S1a&b, the substrate surface after annealing is flat with step-terrace structures of which step height is ~200 pm. The sharp 1×1 and clear Kikuchi patterns, found by in-situ reflection high-energy electron diffraction (RHEED), display the high quality of the Al$_2$O$_3$ (0001) substrate. The AFM image of the surface topology of 30 nm Bi$_2$Te$_3$ (Fig. S1c) shows triangular topography in the same orientation with around 1 nm height steps, corresponding to the thickness of 1 QL Bi$_2$Te$_3$. With the incident electron beam along $[10\bar{1}0]$, the sharp streaky 1×1 pattern of Bi$_2$Te$_3$ [Fig. S1(d)] suggests the good epitaxy and atomically fat surface morphology of Bi$_2$Te$_3$.

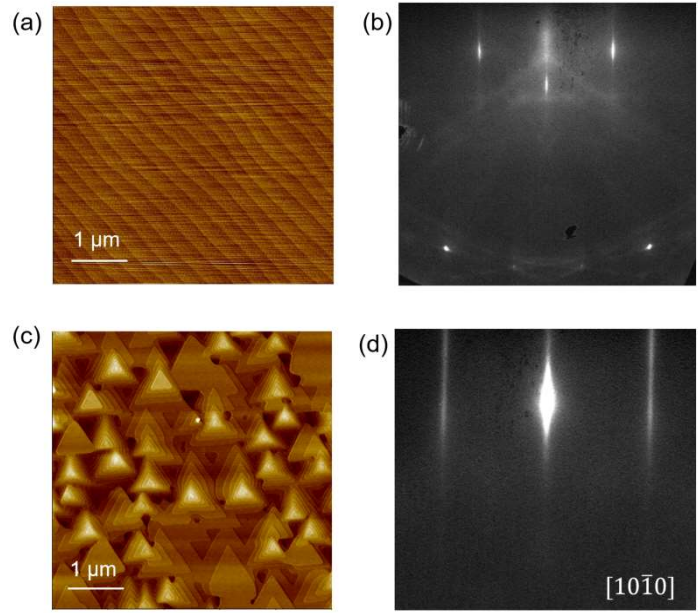

FIG. S1. The characterization of Bi$_2$Te$_3$. (a),(b) AFM image and RHEED pattern of the Al$_2$O$_3$ (0001) substrate, respectively. (c),(d) AFM image and RHEED pattern of Bi$_2$Te$_3$/ Al$_2$O$_3$ (0001) respectively.

We also measured the θ-2θ X-ray diffraction spectra (XRD) of Bi$_2$Te$_3$ films (Fig. S1a) and drew the full width at half maximum (FWHM) of (0015) peak with varying thicknesses by using Lorentz fitting (Fig. S1b). The coherent scattering length ($D_z$) along the growth direction can be estimated by Scherrer's formula, $D_z = k\lambda/(B cos\theta)$, where k is typically set to 0.9 and $B$ is FWHM. As shown in Fig. S1(c), $D_z$ increases as thicknesses increases, revealing improved crystalline quality.

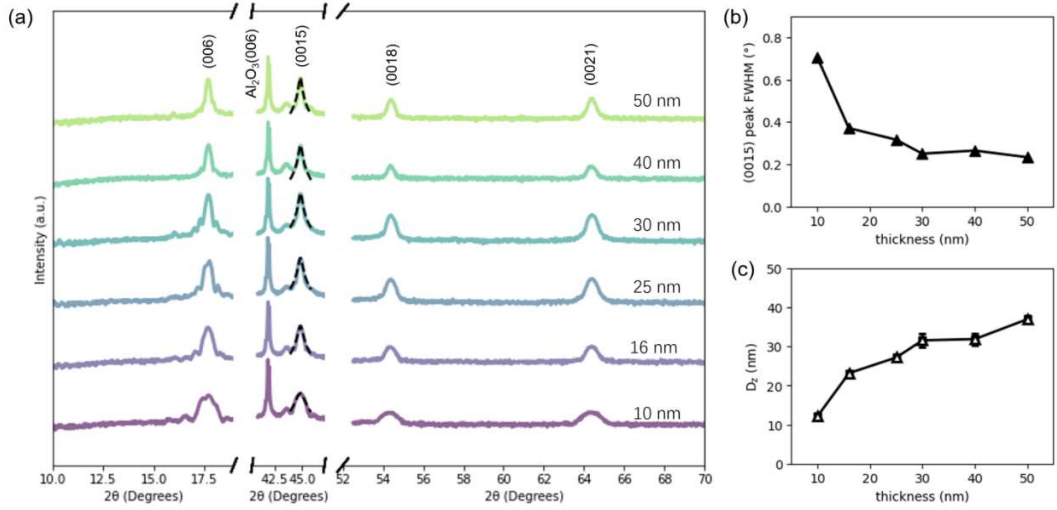

FIG. S2. The XRD characterization of $Bi_2Te_3$. (a) XRD spectra of different thick $Bi_2Te_3$. The dashed line represents the Lorentz fitting curve. (b) The FWHM of the $Bi_2Te_3$ (0015) peak as a function of thickness. (c) Calculated coherent scattering length $D_z$ with different $Bi_2Te_3$ thicknesses.

**2.The calculation of the carrier density and mobility of 2D surface**

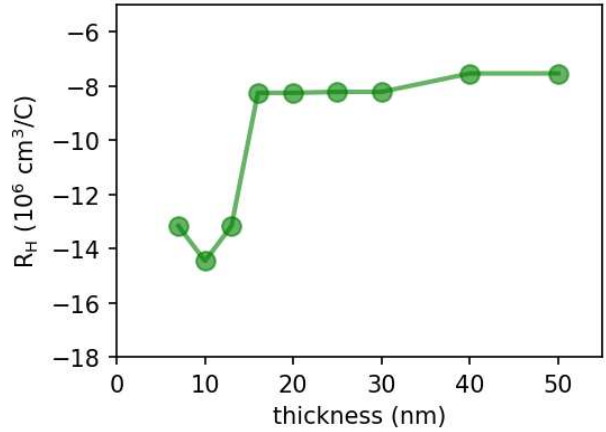

FIG. S3. The thickness dependence of Hall coefficient $R_H$.

For topological insulator, the 2D surface is dominated only if the bulk state is not embedded into the Fermi surface. The 2D carrier density (n) [1] is extracted from $n_{2D} = \frac{1}{eR_H}$, where $R_H = \frac{R_{yx}}{H}$ is low-field Hall coefficient which could be calculated from Fig. 2(b) by measuring resistance with magnetic field changing and $e$ is electronic charge. Based on the variation of $R_H$ with thickness (plotted in Fig.S3), it is evident that when the thickness exceeds 16 nm, $R_H$ exhibits minimal change. The mobility is calculated using $\mu_{2D} = \frac{n_{2D}}{eR_S}$ at zero magnetic field, as plotted in Fig.3(c).

## 3. The second-order nonlinear harmonic response of Bi$_2$Te$_3$ device

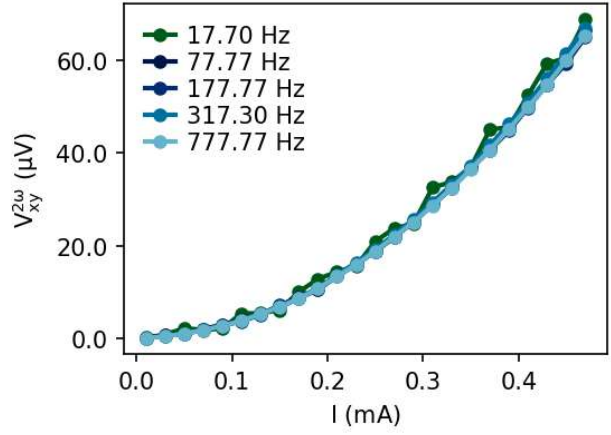

FIG. S4. Frequency dependence. The rectified nonlinear transverse voltage of 7 nm thick Bi$_2$Te$_3$ device for different frequencies, with current along the low symmetry direction at room temperature.

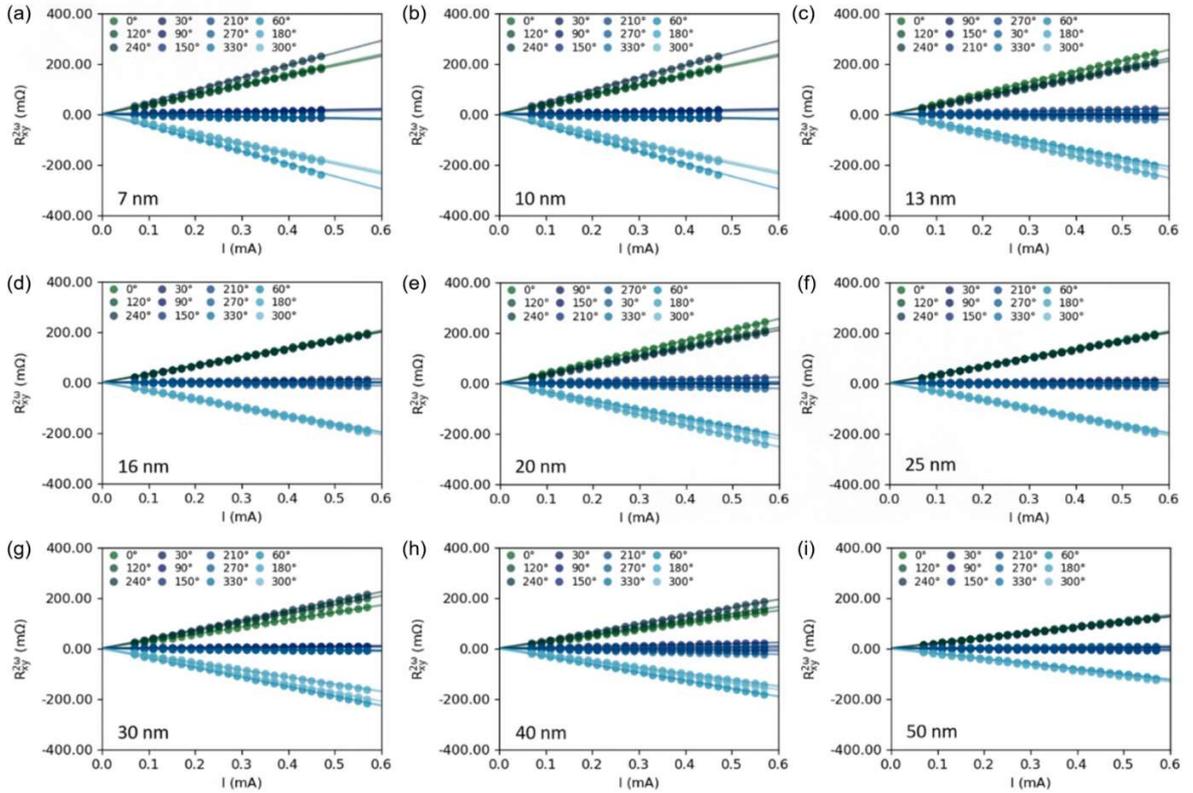

FIG. S5. The nonlinear transport data of Bi$_2$Te$_3$ device. (a-i) $R_{xy}^{2\omega}$ as a function of the input current amplitude $I$ for different thicknesses.

As displayed in Fig. S3, the second harmonic transverse voltage and the input ac current amplitude $I$ exhibit a quadratic relationship. Though the frequency of $I$ varies in the range of 17.7-777.77 Hz, the curve nearly remains unchanged, which reveals the frequency independence of the second-order nonlinear response. For different thicknesses, the second harmonic transverse resistance $R_{xy}^{2\omega}$ is linear to $I$ with different flowing directions as drawn in Fig. S5.